\documentclass[12pt,preprint]{aastex}
\newcommand{\beq}{\begin{equation}}
\newcommand{\beqa}{\begin{eqnarray}}
\newcommand{\eeq}{\end{equation}}
\newcommand{\eeqa}{\end{eqnarray}}
\newcommand{\simg}{\gtrsim}
\newcommand{\siml}{\lesssim}
\newcommand{\meszaros}{M${\acute {\rm e}}$sz${\acute {\rm a}}$ros}
\shorttitle{GeV-TeV emission around GRB remnants}
\shortauthors{Ioka, Kobayashi, \& \meszaros}
\begin{document}
\title{
Extended GeV-TeV Emission around Gamma-Ray Burst Remnants, 
and the Case of W49B
}
\author{
Kunihito Ioka\footnote{
Physics Department and Center for Gravitational Wave Physics, 
104 Davey Laboratory, Pennsylvania State University, University Park, PA 16802},
Shiho Kobayashi$^{1,}$\footnote{
Department of Astronomy and Astrophysics, 525 Davey Laboratory,
Pennsylvania State University, University Park, PA 16802},
and Peter \meszaros$^{1,2,}$\footnote{
The Institute for Advanced Study, Einstein Drive, 
Princeton, NJ 08540}
}

\begin{abstract}
We investigate the high-energy photon emission around Gamma-Ray Burst 
(GRB) remnants caused by ultrahigh-energy cosmic rays (UHECRs) from the
GRBs. We apply the results to the recent report that the supernova 
remnant W49B is a GRB remnant in our Galaxy. If this is correct, 
and if GRBs are sources of UHECRs, a natural consequence of this 
identification would be a detectable TeV photon emission around the 
GRB remnant.  The imaging of the surrounding emission could provide 
new constraints on the jet structure of the GRB.
\end{abstract}


\section{Introduction}

Past Gamma-Ray Burst (GRB) occurrences in our galaxy may have left radio 
remnants resembling hyperenergetic supernovae \citep[e.g.,][]{pernaloeb:00}. 
Recently, Chandra X-ray observations have led to infer that the 
supernova remnant W49B (G43.3-0.2) may be the first remnant of a 
GRB discovered in the Milky Way.\footnote{
http://chandra.harvard.edu/press/04\_releases/press\_060204.html}
This remnant is located at $d\sim 10$ kpc from earth, has a radius of 
$R \sim 10$ pc, and is estimated to have occurred $\sim 3000$ yr ago 
in our Galaxy.

In this Letter we discuss the high energy implications of the GRB
origin of some supernova-like remnants, and in particular of the
assumption that W49B is a GRB remnant. We show that, if we also 
assume that GRBs are sources of ultrahigh-energy cosmic rays (UHECRs),
significant TeV gamma-ray emission may be detectable around W49B.
A positive detection would verify not only that W49B is a GRB remnant,
but also that the GRBs can efficiently accelerate UHECRs.

GRBs are one of the most promising candidates for the origin of UHECRs
\citep{Waxman:1995,Vietri:1995,Milgrom:1995}. The physical conditions 
required in GRBs to produce MeV gamma-rays also allow protons to be
accelerated up to $\sim 10^{20}$ eV. The energy generation 
rate per decade of UHECRs in the range $10^{19}$-$10^{21}$ eV
is similar to the $\gamma$-ray generation rate of GRBs in the BATSE 
energy band $0.02$-$2$ MeV, 
\beqa
E_p^2 \frac{d\dot n_p^{\rm CR}}{dE_p}~(10^{19}-10^{21} {\rm eV}) \simeq
 {\dot \varepsilon}_\gamma^{\rm GRB} (0.02-2 {\rm MeV})~,
\label{eq:edot}
\eeqa
and equation (\ref{eq:edot}) is the common assumption made for
GRBs as the sources of UHECRs \citep{Waxman:2004,Vietri:2003}.
Since photon emission is proportional to the square of the
electron Lorentz factor, two decades in photon energy correspond
to one decade in particle energy, but the electron (and proton) 
energy distribution would be expected to be broader than this. 
Also, additional photon energy is observed in many cases beyond 
the BATSE range. Thus equation 
(\ref{eq:edot}) implying $\dot \varepsilon_{[10^{19},10^{21}] 
{\rm eV}}^{\rm CR} \sim 3 \dot \varepsilon_{\gamma [0.02,2] 
{\rm MeV}}^{\rm GRB}$ reflects approximate equipartition between 
nucleon and electron energies; and larger nucleon energies relative 
to electrons would not be unreasonable \citep[e.g.,][]{Wick:2004}. 
Note that these energies refer to the prompt component, whereas 
snapshot fit and radio calorimetric energies refer to the afterglow.
We note also  that, for the purposes of this Letter, even UHECR of 
energies well below $10^{20}$ eV can lead to GeV-TeV photon emission.

If the GRBs are the origin of UHECRs, an unavoidable component 
of the UHECR outflow will be in the form of neutrons. This is due to
conversion of protons into neutrons via photomeson interactions, 
$p \gamma \to n \pi^{+}$ on the GRB photons. The optical depth for this 
process is about a few for protons with energy $\simg 10^{16}$ eV 
\citep{Waxman:1997,Vietri:1998}. Protons, unlike neutrons, feel the 
magnetic field in the ejecta and are subjected to adiabatic cooling 
as the ejecta expands \citep{Rachen:1998}, hence the neutron to proton
ratio at similar energies could range from approximate equality to 
about 20\% \citep{Atoyan:2001}.
These high energy neutrons decay into protons and electrons 
outside the ejecta, $n \to p + e^{-} + \bar \nu_{e}$, followed by 
interactions with the surroundings. This results in a photon, neutrino 
and neutron emission around the remnant \citep{Dermer:2002,Biermann:2004}.

In this Letter we consider the photon emission from $\beta$-decay 
electrons, and discuss its detectability in particular for the W49B case.
The TeV emission is probably detectable by the future atmospheric 
Cherenkov telescope VERITAS, while HEGRA may have already detected 
this signal in their archival data, if the GRB was intense, and we 
strongly urge the reanalysis of the HEGRA database. 

\section{$\beta$-decay electron emission around the GRB remnant}

We first consider each decade of neutron energy, to simplify the 
discussion. We assume that the GRB ejects neutrons with a Lorentz 
factor $\gamma_{n} \sim 10^{8} \gamma_{n,8}$ and a total energy
$E_{n} \sim 10^{50} E_{n,50}$ erg, at a time $t_{\rm age}\sim 
3000 t_{{\rm age},3.5}$ yr ago (later we integrate over  the energies).
These neutrons decay, $n \to p + e^{-} + \bar \nu_{e}$, over
\beqa
t_{\rm decay} \sim \gamma_{n} t_{\beta} 
\sim 3 \times 10^{3} \gamma_{n,8}\ {\rm yr},
\label{eq:decay}
\eeqa
producing high energy electrons with a Lorentz factor $\gamma_{e} 
\sim \gamma_{n}$. Here $t_{\beta} \sim 900$ s is the $\beta$-decay time 
in the comoving frame. The energy of the $\beta$-decay electrons 
is $m_{p}/m_{e} \sim 10^{3}$ times smaller than that of the neutrons.
We consider only electrons with $\gamma_{e} \sim \gamma_{n} \simg 10^{6}$, 
from neutrons which decay outside the remnant, $c t_{\rm decay} \simg R$,
so that we can easily separate the $\beta$-decay emission
from the remnant emission, such as due to the pion decay \citep{Enomoto:2002}.

The $\beta$-decay electrons radiate via synchrotron and the inverse 
Compton (IC) emission. 
The corresponding fluxes depend on the energy 
densities of the magnetic and photon fields. 
We adopt the typical 
magnetic field in our Galaxy $B \sim 3 \mu$G corresponding to a 
magnetic energy density $U_{B} \sim 4 \times 10^{-13}$ erg cm$^{-3}$.
For the IC process, we consider the cosmic microwave background (CMB) 
as the main target photons.\footnote{
Our calculations show that the IR and optical backgrounds
with mean values are negligible.}
The energy density and temperature of the CMB are about 
$U_{\rm CMB} \sim 4 \times 10^{-13}$ erg cm$^{-3}$
and $\epsilon_{\rm CMB} \sim 2.7$ K.
The electron cooling time due to the combined emissions is 
\beqa
t_{\rm cool}(\gamma_{e}) \sim 
\frac{3 m_{e} c}{4 \gamma_{e} \sigma_{T} U_{\rm total}}
\sim 1 \times 10^{4} \gamma_{e,8}^{-1} U_{{\rm total},-12}^{-1}\ {\rm yr},
\eeqa
where $U_{\rm total}=U_{B}+
\bar U_{\rm CMB}=10^{-12} U_{{\rm total},-12}$ erg cm$^{-3}$ 
is the total energy density, and we use $\bar U_{\rm CMB}
=U_{\rm CMB} \min(1, m_{e} c^{2}/9 \epsilon_{\rm CMB} \gamma_{e})$
to take the Klein-Nishina (KN) effect into account.\footnote{
We numerically found that
the IC spectrum $\nu F_{\nu}$ peaks at $\sim 9 
\epsilon_{\rm CMB} \gamma_{e}^{2}$ in the Thomson regime.}

There are two cases, depending on whether or not most of the neutrons decay 
within the remnant age, the latter case being divided into two sub-cases 
depending on whether or not the electrons cool slower than the remnant age,
leading to three different types of emission region shape.

\begin{figure}
\plotone{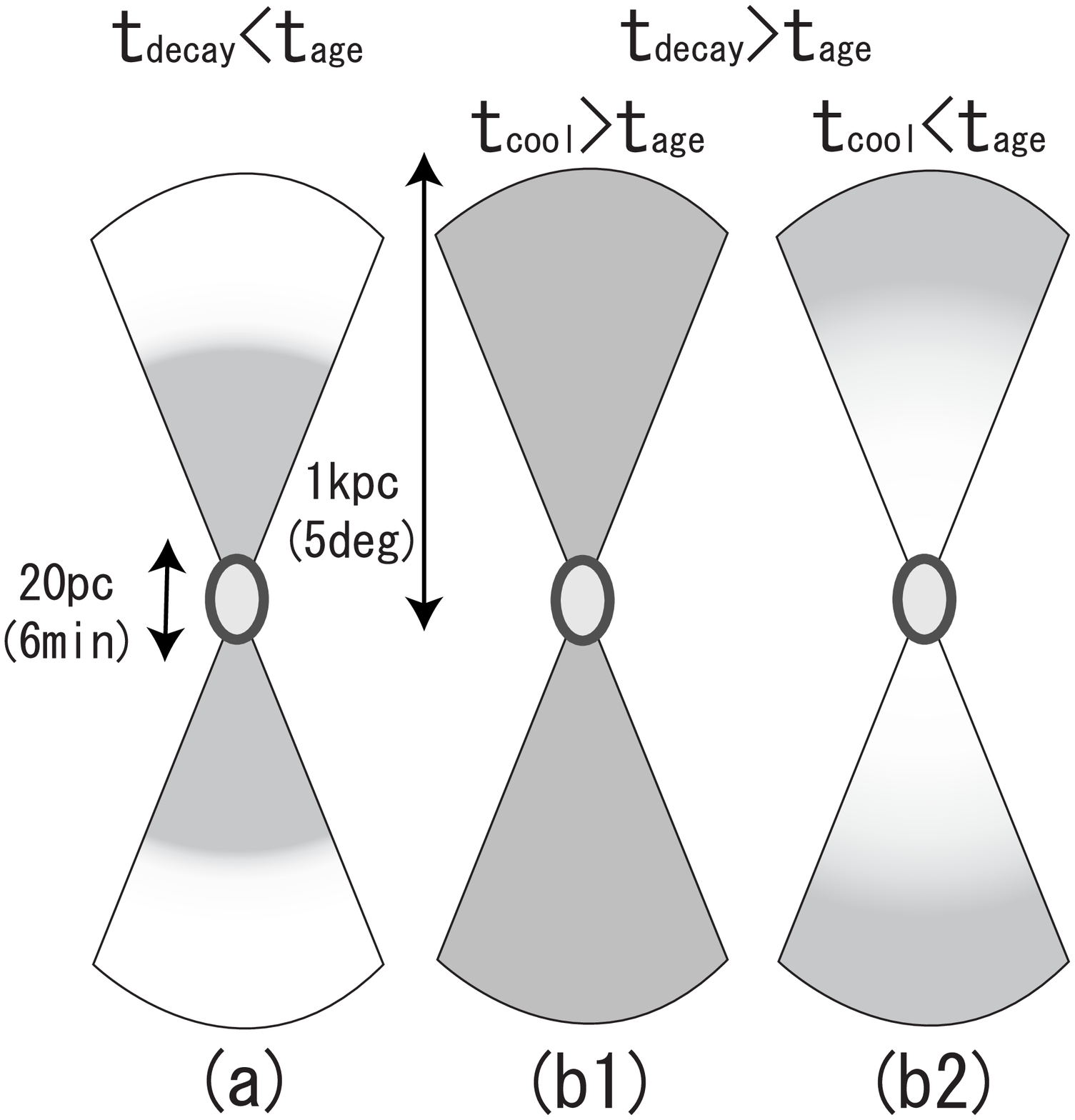}
\caption{\label{fig:image}
The GeV-TeV emission region (shaded region)
on the sky is shown schematically.
(a) The $\beta$-decay time is shorter than the remnant age 
$t_{\rm decay}(\gamma_{e})<t_{\rm age}$, so that
the radius of the emitting region is about the $\beta$-decay length
in equation (\ref{eq:decay}).
The surface brightness on the sky is nearly homogeneous.
(b1) $t_{\rm decay}(\gamma_{e})>t_{\rm age}$
and the initial cooling time is longer than the remnant age
$t_{\rm cool}(\gamma_{e})>t_{\rm age}$.
The radius of the emitting region is about 
$\sim c t_{\rm age}\sim 1 t_{{\rm age},3.5}$ kpc.
The surface brightness is nearly homogeneous.
(b2) $t_{\rm decay}(\gamma_{e})>t_{\rm age}$ and 
$t_{\rm cool}(\gamma_{e})<t_{\rm age}$.
The radius of the emitting region is about 
$\sim c t_{\rm age}\sim 1 t_{{\rm age},3.5}$ kpc.
The jet head region, of size $\sim c t_{\rm cool}(\gamma_{e})$,
has a flux $\sim t_{\rm age}/t_{\rm cool}$ times larger than the rest.
The GRB remnant W49B, shown in the center, has a radius 
$\sim 10$ pc.
}
\end{figure}

(a) Fast decay, $t_{\rm decay}(\gamma_{e})<t_{\rm age}$: In this case
the fraction of neutrons that have decayed is $f_{\beta} \sim 1$.
The radius of the emitting region is roughly equal to the
$\beta$-decay length 
$\sim c t_{\rm decay} \sim 0.9 \gamma_{n,8}$ kpc,
which is much larger than 
the Larmor radius of the $\beta$-decay electrons,
\beqa
r_{L} \sim \frac{\gamma_{e} m_{e} c^{2}}{q B}
\sim 6 \times 10^{-2} \gamma_{e,8} B_{-6}^{-1}\ {\rm pc},
\eeqa
where $B=10^{-6} B_{-6}$ G. Thus, 
the sky distribution of the emitting 
region is roughly preserved, as shown in Figure~\ref{fig:image}~(a).
The emitting region has an elongated shape on the sky,
whose solid angle is approximately
\beqa
\Omega \sim 4 \theta (c t_{\rm decay}/d)^{2}
\sim (1 \theta_{-1} \gamma_{e,8})^{\circ} \times (10 \gamma_{e,8})^{\circ}
\sim 3 \times 10^{-3} \theta_{-1} \gamma_{e,8}^{2}\ {\rm sr},
\eeqa
if the opening half-angle of the jet is $\theta=0.1 \theta_{-1}$.
Note that the neutrons are beamed into a jet because of 
the relativistic beaming.

An electron with an initial cooling time $t_{\rm cool}(\gamma_{e})<t_{\rm age}$
cools down to the Lorentz factor determined by 
$t_{\rm cool}(\gamma_{e})\sim t_{\rm age}$.
Thus the current Lorentz factor of electrons is given by
\beqa
\hat \gamma_{e} \sim 
\min \left(\gamma_{e}, \frac{3 m_{e} c}{4 t_{\rm age} \sigma_{T} U_{\rm total}}
\right)
\sim 1\times 10^{8} 
\min(\gamma_{e,8}, 3 t_{{\rm age},3.5}^{-1} U_{{\rm total},-12}^{-1}).
\eeqa
Irrespective of the initial $\gamma_{e}$, the order of magnitude of the 
current Lorentz factor $\hat \gamma_{e}$ is the same in all the emitting 
region, so that the surface brightness on the sky depends weakly on the 
radius $\propto r^{-1}$, and it is nearly homogeneous.

The characteristic synchrotron frequency
and the synchrotron peak flux at the observer are given by
\beqa
\nu_{m}&=&\frac{q B \hat \gamma_{e}^{2}}{2 \pi m_{e} c}
\sim 1 \times 10^{2} B_{-6} \hat \gamma_{e,8}^{2}\ {\rm eV},
\nonumber\\
\nu F_{\nu}(\nu_{m}) &\sim&
\frac{m_{e}}{m_{p}} \frac{\hat \gamma_{e}}{\gamma_{e}}
\frac{U_{B}}{U_{\rm total}}
\frac{f_{\beta} E_{n}}{4 \pi d^{2} t_{\rm cool}(\hat \gamma_{e})}
\sim 1 \times 10^{-11} f_{\beta} E_{n,50} U_{B,-12}
\hat \gamma_{e,8}^{2} \gamma_{e,8}^{-1}\
{\rm erg}\ {\rm s}^{-1}\ {\rm cm}^{-2}.
\label{eq:syn}
\eeqa
Similarly the characteristic IC frequency
and the IC peak flux are
\beqa
\nu_{\rm CMB} &\sim& \min(\hat \gamma_{e} m_{e} c^{2}, 
9 \epsilon_{\rm CMB} \hat \gamma_{e}^{2})
\sim 50 \min(\hat \gamma_{e,8}, 
2 \epsilon_{{\rm CMB},-3} \hat \gamma_{e,8}^{2})\ {\rm TeV},
\nonumber\\
\nu F_{\nu}(\nu_{\rm CMB}) &\sim&
\frac{m_{e}}{m_{p}} \frac{\hat \gamma_{e}}{\gamma_{e}}
\frac{\bar U_{\rm CMB}}{U_{\rm total}}
\frac{f_{\beta} E_{n}}{4 \pi d^{2} t_{\rm cool}(\hat \gamma_{e})}
\sim 1 \times 10^{-11} f_{\beta} E_{n,50} \bar U_{{\rm CMB},-12}
\hat \gamma_{e,8}^{2} \gamma_{e,8}^{-1}\
{\rm erg}\ {\rm s}^{-1}\ {\rm cm}^{-2},
\label{eq:IC}
\eeqa
where $\epsilon_{\rm CMB}=10^{-3} \epsilon_{{\rm CMB},-3}$ eV
is the target photon energy.
The surface brightness of emission is about $S=\nu F_{\nu}/\Omega$ 
$\sim 5 \times 10^{-9} \theta_{-1}^{-1} E_{n,50} U_{-12}
\hat \gamma_{e,8}^{2} \gamma_{e,8}^{-3}$
erg s$^{-1}$ cm$^{-2}$ sr$^{-1}$.

(b) Slow decay, $t_{\rm decay}(\gamma_{e})>t_{\rm age}$. In this case,
almost all neutrons are still decaying. 
The radius of the emitting region is roughly
$\sim c t_{\rm age}\sim 1 t_{{\rm age},3.5}$ kpc, which is again 
larger than the Larmor radius for our parameters.
In this case, however, the shape of the emitting region depends on 
whether electrons cool within the remnant age or not.

(b1) If $t_{\rm cool}(\gamma_{e})>t_{\rm age}$, the electrons do not 
cool much, so the current Lorentz factor of the electrons is 
$\hat \gamma_{e}\sim \gamma_{e}$.  Thus the surface brightness of 
the emitting region is nearly homogeneous 
(see Figure~\ref{fig:image}~(b1)).
The fraction of neutrons that have decayed is 
$f_{\beta} \sim t_{\rm age}/t_{\rm decay} \sim 1 t_{{\rm age},3.5}
\gamma_{e,8}^{-1}$.
The solid angle of the emitting region on the sky is about
\beqa
\Omega \sim 4 \theta (c t_{\rm age}/d)^{2} 
\sim (1 \theta_{-1} t_{{\rm age},3.5})^{\circ} 
\times (10 t_{{\rm age},3.5})^{\circ}
\sim 4 \times 10^{-3} \theta_{-1} t_{{\rm age},3.5}^{2}\ {\rm sr}.
\eeqa
The synchrotron and IC emission is given by equations (\ref{eq:syn})
and (\ref{eq:IC}), respectively.
The surface brightness of the emission is 
$S=\nu F_{\nu}/\Omega$ 
$\sim 4 \times 10^{-9} \theta_{-1}^{-1} E_{n,50} U_{-12}
t_{{\rm age},3.5}^{-1}$
erg s$^{-1}$ cm$^{-2}$ sr$^{-1}$.

(b2) If $t_{\rm cool}(\gamma_{e})<t_{\rm age}$, almost all the 
electrons cool down, except for a smaller fraction of electrons at 
the jet head which have not cooled yet because the neutrons are 
just decaying, $t_{\rm decay}(\gamma_{e}) > t_{\rm age}$.
Although the number of these hot electrons is small,
the flux from these hot electrons dominates that of the rest,
since they have a larger energy and a shorter cooling time 
(see Figure~\ref{fig:image}~(b2)).
Thus we concentrate on the hot electrons within the distance 
$\sim c t_{\rm cool}(\gamma_{e}) 
\sim 3 \gamma_{e,8}^{-1} U_{\rm total,-12}^{-1}$ kpc 
from the jet head.
The current Lorentz factor of these electrons is 
$\hat \gamma_{e} \sim \gamma_{e}$, and
the fraction of neutrons that have decayed is 
$f_{\beta} \sim t_{\rm cool}/t_{\rm decay} \sim 
3 \gamma_{e,8}^{-2} U_{\rm total,-12}^{-1}$.
The solid angle of the emitting region on the sky is 
\beqa
\Omega &\sim& 4 \theta (c t_{\rm age}/d) (c t_{\rm cool}/d)
\sim (1 \theta_{-1} t_{{\rm age},3.5})^{\circ} \times
(30 \gamma_{e,8}^{-1} U_{{\rm total},-12}^{-1})^{\circ}
\nonumber\\
&\sim& 1 \times 10^{-2} \theta_{-1} t_{{\rm age},3.5} \gamma_{e,8}^{-1} 
U_{{\rm total},-12}^{-1}\ {\rm sr}.
\eeqa
The synchrotron and IC emission is given by equations (\ref{eq:syn})
and (\ref{eq:IC}), respectively.
The surface brightness of the emission is 
about $S=\nu F_{\nu}/\Omega$ 
$\sim 1 \times 10^{-9} \theta_{-1}^{-1} E_{n,50} U_{-12}
t_{{\rm age},3.5}^{-1}$
erg s$^{-1}$ cm$^{-2}$ sr$^{-1}$.

We may usually assume the emission to be isotropic in the lab-frame,
since the Larmor radius is the minimum scale 
$r_{L}<\min(c t_{\rm cool},c t_{\rm age})$ for almost all parameters.
For $\gamma_{e} \simg 10^{10} 
\min(2 B_{-6}^{1/2} U_{{\rm total},-12}^{-1/2},
10^{2} B_{-6} t_{{\rm age},3.5})$, 
however, the emission is beamed into the jet direction and
the observed flux is reduced.

\section{Detectability}

We have numerically calculated the IC spectrum using equation (2.48)
of \citet{Blumenthal:1970}. 
We assume a blackbody CMB spectrum and a power-law neutron distribution
$N(\gamma_{n}) d\gamma_{n} \propto \gamma_{n}^{-p} d\gamma_{n}$ with $p=2$ 
for $10^{7} \le \gamma_{n} < 10^{11}$ and $p=1$ for $\gamma_{n} < 10^{7}$ 
as expected in GRBs \citep{Waxman:1997}, and use $f_{\beta} N(\gamma_{e}
<\hat \gamma_{e}) d\gamma_{e}$ as the distribution of emitting electrons,
where $f_{\beta}$ includes the effects of the cooling and the partial
neutron decay. We consider three cases for the normalization of $N(\gamma_{e})$:
(I) We set it equal to the geometrically corrected energy of 
equation (\ref{eq:edot}), $\int_{10^{10}}^{10^{12}} \gamma_{e} m_{n} c^{2} 
N(\gamma_{e}) d\gamma_{e}=3 \times 10^{51}$ erg.
(II) The same total energy is distributed over the entire 
range, $\int_{1}^{10^{11}} \gamma_{e} m_{n} c^{2} 
N(\gamma_{e}) d\gamma_{e}=3 \times 10^{51}$ erg.
(III) The energy of case (II) is further reduced by a factor $5$,
considering that only $\sim 20\%$ of accelerated protons are converted 
to neutrons \citep{Atoyan:2001,Dermer:2004}.

The emission from the $\beta$-decay electrons is extended,
hence it competes with the diffuse background radiation.
Since W49B resides in the Galactic disk $(l,b)=(43.3,-0.2)$,
the disk emission is the main confusion source.
We find that the IR-optical band \citep[e.g.,][]{Bernstein:2002} and the 
X-ray band \citep{Kaneda:1997,Snowden:1995} are not suitable for the 
detection of the $\beta$-decay emission considered here, because the 
diffuse background dominates.
The MeV region \citep{Kinzer:1999,Strong:2000} could be a possible 
observing window, but appropriate detectors have not been developed 
in this band.  Although INTEGRAL might be one possibility, the sub-MeV 
background may be difficult to subtract \citep{Lebrun:2004}.

On the other hand, the GeV band \citep{Hunter:1997} and the TeV band 
appear to be suitable windows for the observation of $\beta$-decay IC
emission. Strictly speaking, the diffuse background in the TeV band
has not been detected. Only an upper limit exists 
\citep{Aharonian:2001,LeBohec:2000,Amenomori:2002}.
However the predicted background is below the power-law extrapolation 
of the upper limit \citep{Aharonian:2000,Strong:2004},
so that it is probable that the TeV background is negligible.

\begin{figure}
\plotone{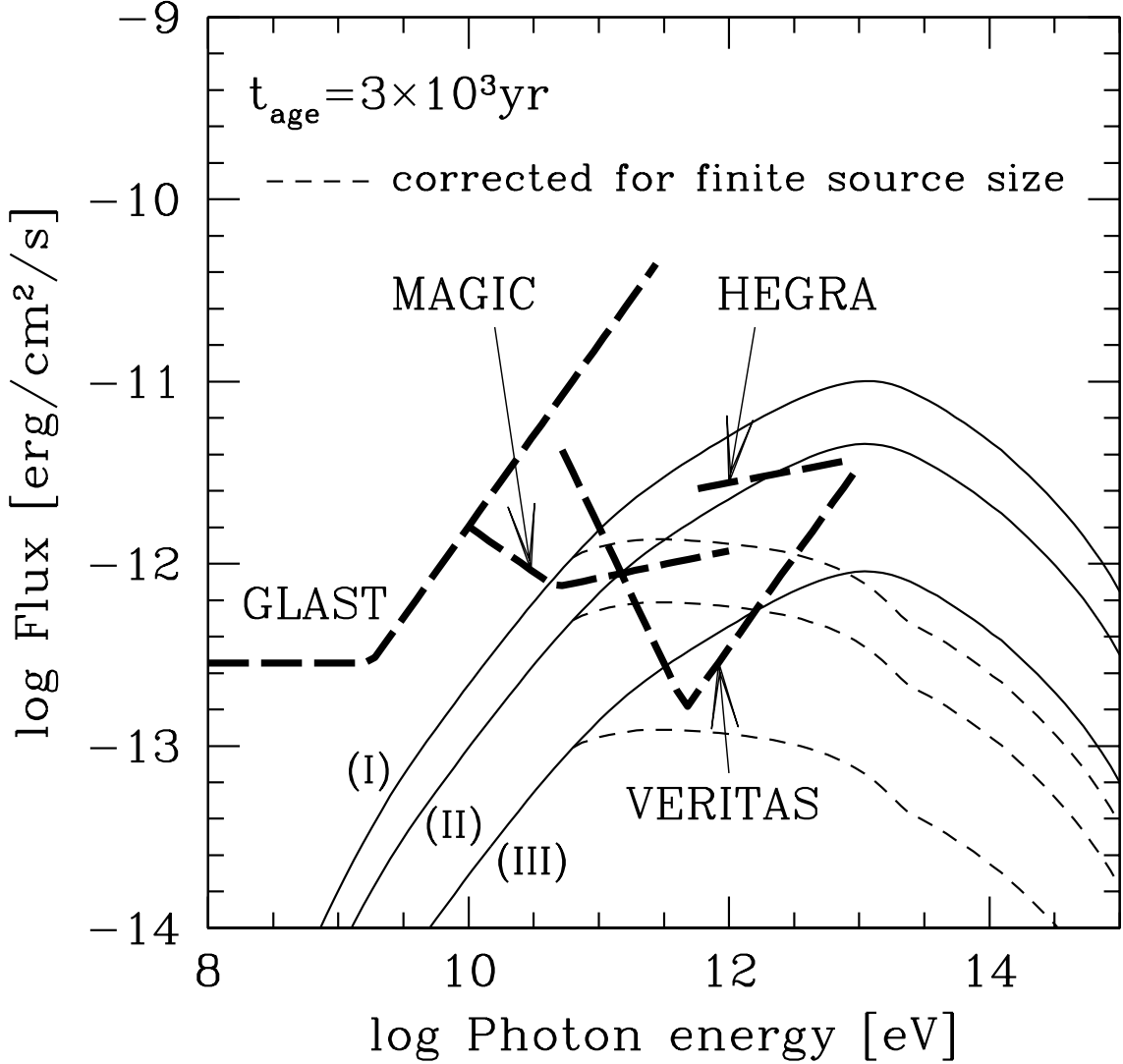}
\caption{\label{fig:flux}
The flux of IC emission from the $\beta$-decay electrons 
(solid lines) is compared with the detector sensitivity 
(bold long dashed lines).
Three cases for the total energy are plotted:
(I) Neutrons in the range $10^{10}<\gamma_{n}<10^{12}$ have 
the geometrically corrected energy $3\times 10^{51}$ erg
according to equation (\ref{eq:edot}).
(II) The energy $3\times 10^{51}$ erg is spread among all neutrons 
with $1<\gamma_n <10^{11}$.
(III) The energy of case (II) is further reduced by a factor $5$.
The dashed lines are the flux of $\beta$-decay emission
multiplied by $(\Omega/\pi \theta_{\rm cut})^{-1/2}$
in order to take the finite source size into account,
where $\Omega$ is the solid angle of the emitting region on the sky
and $\theta_{\rm cut}\sim 0.1^{\circ}$ is the angular cut in the analysis.
The sensitivities of HEGRA, MAGIC and VERITAS should be compared 
with the dashed lines.  The remnant age is $t_{\rm age}=
3\times 10^{3}$ yr, and the distance is $d=10$ kpc.
}
\end{figure}

We compare now the predicted GeV-TeV flux to the sensitivities of
various detectors. 
Figure~\ref{fig:flux} shows the flux of the 
$\beta$-decay electrons and the sensitivities of GLAST, HEGRA, MAGIC\footnote{
http://hegra1.mppmu.mpg.de/MAGICWeb/} and 
VERITAS \citep[e.g.,][]{McEnery:2004}. Note that the northern sky 
location of W49B makes it an unsuitable target for some ground-based detectors,
such as CANGAROO or HESS.

With HEGRA, MAGIC and VERITAS, whose angular resolution is $\sim 
0.1^{\circ}$, the source size should be taken into account.  Since the 
search region has to be expanded, more background is included.  
The sensitivity is proportional to the inverse square of the background, 
for background dominated counting statistics.  Therefore the flux 
sensitivity of an atmospheric Cherenkov telescope to an extended source, 
$F_{\nu}^{\rm extend}$, is given by $F_{\nu}^{\rm extend}=
F_{\nu}^{\rm point} (\Omega/\pi \theta_{\rm cut}^{2})^{1/2}$,
where $F_{\nu}^{\rm extend}$ is the sensitivity to a point source,
and $\theta_{\rm cut} \sim 0.1^{\circ}$ is the angular cut in the 
point source analysis \citep{Konopelko:2002,Lessard:2001}.
In Figure~\ref{fig:flux} the dashed lines show the corrected 
$\beta$-decay emission
$\nu F_{\nu} (\Omega/\pi \theta_{\rm cut}^{2})^{-1/2}$,
which should be compared with the detector sensitivity. 

From Figure~\ref{fig:flux} we see that VERITAS might be likely to detect 
the $\beta$-decay emission. Even our most conservative case (III) is only 
a factor $2$ below the VERITAS limit. MAGIC may also marginally detect
the emission. GLAST appears unable to detect the 
signal, while the sensitivity of HEGRA is only a factor $2$ 
above the flux of case (I). Thus, in the case of a stronger than average
GRB \citep{Wick:2004}, the detection by HEGRA may not be implausible.
In fact the observed total gamma-ray energy has a factor $2$ dispersion
\citep{Bloom:2003}.

HEGRA has actually observed the region including W49B
\citep{Aharonian:2001,Aharonian:2002}. An upper limit on the TeV flux 
from W49B is given as $\sim 0.14$ Crab flux
$\sim 7 \times 10^{-12}$ erg cm$^{-2}$ s$^{-1}$ \citep{Aharonian:2002}.
However this limit constrains the flux only in a $\sim 0.1^{\circ}$ 
circle centered on W49B, since the angular cut $\theta_{\rm cut} \sim 
0.1^{\circ}$ is applied in the analysis. Thus this limit is not stringent.
In order to improve the sensitivity to the $\beta$-decay emission,
we should expand the angular cut $\theta_{\rm cut}$ so that all the 
$\beta$-decay emitting region is included in the analysis.
The probable size is $\sim 0.1^{\circ}\times 1^{\circ}$,
centered on W49B and with a shape as in Figure~\ref{fig:image}~(a), 
possibly along the direction in which more metals are ejected within W49B.
Although HEGRA has ended observations, a reanalysis of the data on 
this region may constrain the $\beta$-decay emission.

\section{Discussion}

The possibility of imaging the $\beta$-decay emission region 
of a GRB remnant could open a novel way of constraining the
structure of GRB jets.  Currently jets are unresolved, but
recent studies indicate that the jet structure is essential 
to an understanding the GRB phenomenon
\citep{Rossi:2002,Zhang:2002,Zhang:2004}.
Depending on the jet structure and the viewing angle,
the same jet may be observed as different phenomena, such as
short GRBs, long GRBs, X-ray flashes and X-ray rich GRBs
\citep{Yamazaki:2004,Yamazaki:2002,Ioka:2001}.
It is difficult to determine the jet structure from
observations of the photon light curve and spectra 
during the GRB and afterglow phase, because the physical
dimensions in these stages are much smaller than the decay 
lengths considered here, and the relativistic beaming prevents 
observing the whole angle of the jet. TeV imaging of the 
$\beta$-decay emission could be a possible way\footnote{Other 
methods might include gravitational waves \citep{Sago:2004,Kobayashi:2003}.}
to determine or constrain the jet structure, since it provides
a much longer lever arm to trace the inner jet. 

There are a number of uncertainties which could affect the conclusions.
For instance, the photon energy density field may be higher than estimated 
from the diffuse backgrounds, due to contributions from the nearby 
H II region W49A. The latter emits about $10^{51}$ Lyman continuum 
photons s$^{-1}$ \citep{Conti:2002}, hence the corresponding photon energy 
density is 
$\sim 10^{-14} (d_{A}/1\ {\rm kpc})^{-2}$ erg cm$^{-3}$, where 
$d_{A}$ is the distance from W49A.
Since the projected distance between W49A and W49B is $\sim 40$ pc,
the IC emission within $\sim 100$ pc of W49B may be enhanced.
Also, if the jet is directed towards us, relativistic effects such 
as a superluminal motion or differential Doppler boosting may be 
important. However, the barrel shape of W49B possibly suggests that 
the jet is not directed towards us.
Nevertheless the forward- and counter-jet sizes may differ by
factors of a few since it takes about the observed age for light to 
cross the system.

To infer the jet structure precisely, the sky distribution of 
the electrons must be nearly preserved.  Thus the diffusion 
length of electrons has to be smaller than the emitting region.
The diffusion length 
$r_{D} \sim (\kappa t_{\rm age})^{1/2} 
\sim 100 \gamma_{e,7}^{1/6} t_{{\rm age},3.5}^{1/2}$ pc 
within the remnant age $t_{\rm age}$ is smaller than 
the emitting region for $\gamma_{e}\sim 10^{7}$,
if we use the diffusion coefficient of electrons
$\kappa \sim 9 \times 10^{28} \gamma_{e,7}^{1/3}$ cm$^{2}$ s$^{-1}$
\citep{Wick:2004}.
Electrons with $\gamma_{e} \siml 10^{7}$ may suffer diffusion,
while for $\gamma_{e} \simg 10^{9}$ the emission is in the regime (b2) 
and the diffusion may increase the cooling region size by at most 
a factor $2$.
However we should note that the diffusion 
coefficient has large uncertainties.

Even if W49B is not a GRB remnant, there may exist other, older 
GRB remnants, whose age may be $\lesssim 10^{5}$ yr, since the 
(collimation corrected) GRB rate is $ \gtrsim 10^{-5}$ yr$^{-1}$ 
galaxy$^{-1}$.  
The same formalism may be used to discuss the detectability.
As long as the electron diffusion length does not become  a limiting 
factor, the $\beta$-decay emission from such an older remnant may be 
detectable.

\acknowledgments
We thank S.~Razzaque, K.~Mori, S.~Park, J.~Granot, E.~Ramirez-Ruiz
and the referee for useful comments.
This work was supported in part by the Center for Gravitational 
Wave Physics under the National Science Foundation cooperative 
agreement PHY 01-14375 (KI, SK), NASA NAG5-13286, NSF AST 0098416
and the Monell Foundation.

%
%

%
%

\newpage

\newpage

\end{document}